\documentclass[a4paper]{jpconf}
\usepackage{graphicx}
\usepackage{wrapfig}
\usepackage{array,multirow,graphicx}
\usepackage{wrapfig,lipsum,booktabs}
\usepackage{lineno}
\begin{document}
\title{	Search for FCNC in top quark production and decays}

\author{Reza Goldouzian on behalf of the CDF, D0, ATLAS and CMS collaborations}

\address{}

\ead{reza.goldouzian@cern.ch}
\begin{abstract}
An overview of searches for top quark flavour changing neutral currents (FCNC) from the CDF, D0, ATLAS and CMS collaborations is presented. The FCNC interactions of top quarks  are probed through the anomalous decays and production channels. No clear evidence for the presence of the signal is observed in none of the searches and the upper limits are set at 95\% confidence level (CL) on the cross sections, anomalous couplings and the branching ratios.
\end{abstract}

\section{Introduction}
Flavor Chaning Neutral Currents (FCNC) are highly suppressed in the standard model (SM) due to the GIM mechanism \cite{gim}. For this reason, the SM predicts very small rates for the branching ratios of top quark FCNC decays to an up type quark and a neutral gauge or a Higgs bosons: $Br(t\rightarrow X(=\gamma,Z,g,H)+ q(=c,u)) < 10^{-10}$. Many models beyond the SM predict a huge enhancement on the expected rate of top quark FCNC branching ratios by introducing new heavy particles which can contribute into the loops. Any observation of these decays would indicate new physics. 

In order to observe a deviation from the SM predictions independent of the specific new physics scenarios, effective Lagrangian approach is used to search for the top quark FCNC signal in all analysis that will be reviewed. Following reference \cite{smbr}, the most general effective Lagrangian describing the top quark FCNC interactions can be written as follows:
\begin{eqnarray}
\label{lagrangy}
-\mathcal{L}_{eff} & =  \frac{g}{2c_W}X_{qt}\overline{q}\gamma_\mu(x_{qt}^LP_L+x_{qt}^RP_R)tZ^\mu +  \frac{g}{2c_W}X_{qt}\kappa_{qt}\overline{q} (\kappa_{qt}^v+\kappa_{qt}^a\gamma_5)\frac{i\sigma_{\mu\nu}q^\nu}{m_t}tZ_\mu   \\ \nonumber
 & +  e\lambda_{qt}\overline{q}(\lambda_{qt}^v+\lambda_{qt}^a\gamma_5)\frac{i\sigma_{\mu\nu}q^\nu}{m_t}tA^\mu+g_s\zeta_{qt}\overline{q}(\zeta_{tq}^v+\zeta_{qt}^a\gamma_5)\frac{i\sigma_{\mu\nu}q^\nu}{m_t}T^aqG^{a\mu}  & \\ \nonumber
 & +  \frac{g}{2\sqrt{2}}g_{qt}\overline{q}(g_{qt}^v+g_{qt}^a\gamma_5)tH + h.c. &  \nonumber
\end{eqnarray}
The parameters $X_{qt}$, $\kappa_{qt}$, $\lambda_{qt}$, $\zeta_{qt}$, and $g_{qt}$ define the anomalous coupling constants which are normalized as $|x_{qt^L}|^2+|x_{qt}^R|^2=1$, $|\kappa_{qt}^\nu|^2+|\kappa_{qt}^a|^2=1$, etc. Different measurements use different normalization of coupling constants in $\mathcal{L}_{eff}$, making the comparison of the couplings not straightforward. Therefore, the limits on the anomalous couplings are based on the normalization of the coupling constants in their corresponding publications. Results on the branching ratios are more easily comparable.

In the following the latest analyses of top quark FCNC interactions performed at CMS \cite{Chatrchyan:2008aa}, ATLAS \cite{Aad:2008zzm}, CDF \cite{Acosta:2004yw} and D0 \cite{Abazov:2005pn} are summarised.

\section{Searches for t-q-Higgs anomalous interactions}
The discovery of a new boson with a mass around 125 GeV in 2012 at the LHC \cite{Aad:2012higgs,Chatrchyan:2012higgs} compatible with the SM Higgs boson, opens up the possibility of searching for anomalous top quark decays to an up type quark and a Higgs boson.  Considering the relatively large production cross section of $t\bar{t}$ at the LHC, the ATLAS \cite{atlasTQH} and CMS \cite{CMSTQH} collaborations have searched for $t\rightarrow ch$ in $t\bar{t}$ events.

The ATLAS collaboration has chosen the $H\rightarrow \gamma \gamma$ decay mode. Despite its small branching ratio ($\sim$0.23 \%), $H\rightarrow \gamma \gamma$ leads to a clean signature. Signal events corresponding to
$t\bar{t}$ production, with one top quark decaying into a charm quark and a Higgs boson  were generated using PROTOS 2.2 \cite{AguilarSaavedra:2009mx}. Two final states with the hadronic and leptonic decays of W boson from the remaining top quark decay are analysed. 
Events are first required to have two well identified and isolated photon with $E_T$ greater than 40 GeV and 30 GeV for leading and subleading photon candidates. In hadronic channel, events are required to have at least four jets among which at least one is b-tagged. After the selection of four jets, one top-quark candidate is constructed from the two photons and one jet ($m_{\gamma \gamma j}$), and another top-quark candidate is formed from the three remaining jets ($m_{j j j}$) which are required to lie within certain mass windows ($156<m_{\gamma \gamma j}<191$ GeV and $130<m_{j j j}<210$ GeV). In leptonic channel, events with exactly one electron or muon and with two or more jets which at least one is b-tagged are selected. After the selection of two jets, one top quark candidate is constructed from the two photons and one jet ($m_{\gamma \gamma j}$), and another top-quark candidate is built from the remaining selected jet, the lepton and the neutrino ($m_{l \nu j}$) which are required to lie within certain mass windows ($156<m_{\gamma \gamma j}<191$ GeV and $135<m_{l \nu j}<205$ GeV).
The contributions of the SM Higgs production (resonant background), $t\bar{t}$ and W$\gamma$ (non-resonant backgrounds) are estimated from the simulation. Non-resonant production of diphoton final states with several additional jets which is the dominated background in the hadronic selection is normalised to data. Non-resonant backgrounds are small once a $t\bar{t}$-like topology is requested. 

The diphoton mass spectrum is fitted using a likelihood function for the individual search channels. In the hadronic channel, the $m_{\gamma\gamma}$ distribution is fitted using 4.7 fb$^{-1}$ of 7 TeV and 20.3 fb$^{-1}$ of 8 TeV data. In the leptonic channel, the analysis is based on the event counting in two the $m_{\gamma\gamma}$ regions using 8 TeV data. No significant signal is observed and an upper limit at the 95\% CL is set on the $t\rightarrow qH$ branching ratio to be 0.79\%.

The CMS search for $t\rightarrow qH$ is a reinterpretation of an inclusive multilepton analysis \cite{Chatrchyan:2014aea} and a search involving lepton and photon final states  \cite{CMS:2013eua} conducted by CMS collaboration in terms of FCNC parameters.  In addition to diphoton decay mode of the Higgs boson, leptonic final state decay modes are also considered. These include H$\rightarrow$ WW$^*$ $\rightarrow l\nu l \nu$, 
H$\rightarrow\tau \tau$ and H$\rightarrow$ ZZ$^*$ $\rightarrow jjll,\nu \nu ll,llll$. Signal events were generated with MadGraph5 event generator \cite{Maltoni:2002qb}.
Multilepton candidate events are categorized into mutually exclusive search channels to maximize the overall search sensitivity.  Events are classified by the presence of a 
hadronic $\tau$, a b-jet, an opposite-sign same-flavor (OSSF) dilepton pair in which the invariant mass can be inside, above or bellow Z boson mass. The events are further classified by the missing transverse energy and hadronic activities which is measured by the scalar sum of the selected jets. Diphoton candidate events must have one lepton of any flavor and two well isolated photon. The events are subdivided into four E$_T^{miss}$ bins with lower edges at 0, 30, 50, and 100 GeV.

The signal predominantly populates channels that have three leptons (no hadronic $\tau$), a b-jet, and no OSSF pair or an OSSF pair off Z, as well as diphoton channels with a b-tag.
No significant excess is observed. A counting experiment is done with several channels to compute the limit.
The upper limit on Br($t\rightarrow qH$) obtained from the combination of the multilepton channels with diphoton channels is 0.56\%.

\section{Searches for t-q-$\gamma$ anomalous interactions}

The FCNC-induced couplings of the type $tq\gamma$ have been explored at the Tevatron for the first time by looking at the anomalous top decays in $t\bar{t}$ events \cite{Abe:1997fz}. Considering the statistical uncertainty as the main source of systematic uncertainties, the upper limit at 95\% CL is set on the branching fraction Br($t\rightarrow q\gamma$)$<3.2\%$. In $e^+ e^-$ collisions at LEP and $ep$ collisions at HERA, the FCNC $tq\gamma$ have been probed through the single top production with no evidence of any signal \cite{Abramowicz:2011tv,Aaron:2009vv,Abdallah:2003wf}.
\begin{wrapfigure}{r}{0.4\textwidth}
  \begin{center}
    \includegraphics[width=0.4\textwidth]{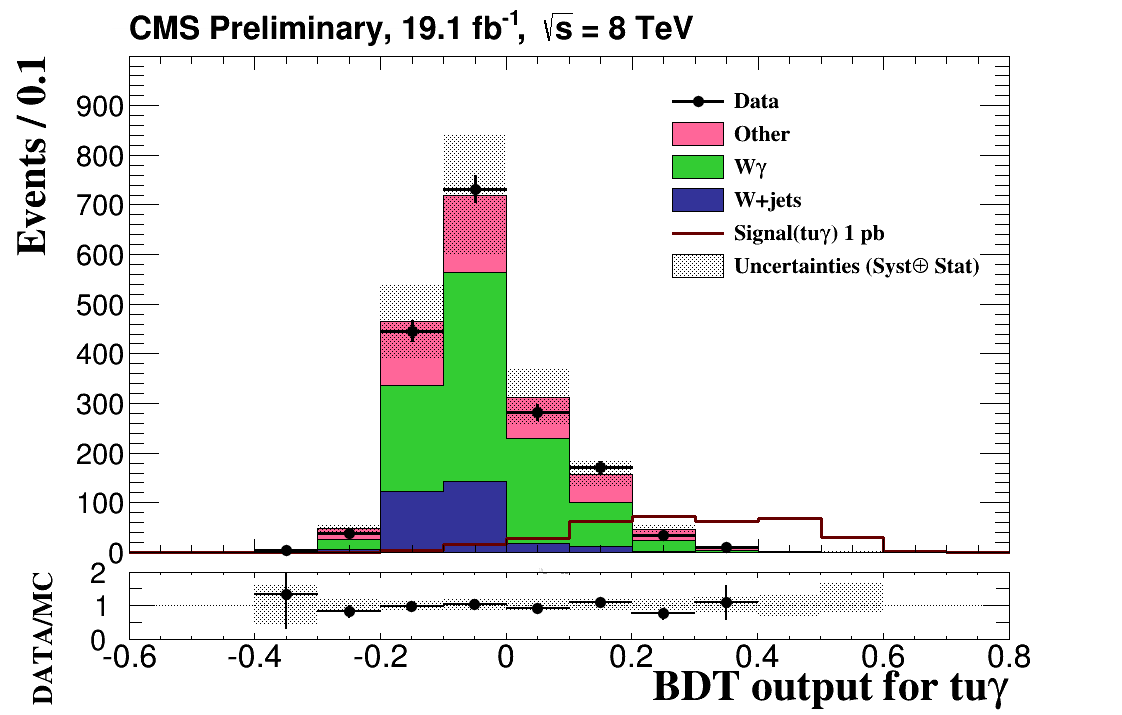}
  \end{center}
  \begin{minipage}[b]{14pc}\caption{\label{tugammaBDT} BDT output distribution for the $tu\gamma$ couplings. The signal is normalized to a cross section of 1 pb.}
\end{minipage}
\end{wrapfigure}
The CMS collaboration conducted a search for FCNC in the production of single top quark in association with a photon \cite{CMS:2014hwa}. Considering the muonic decay of W boson from top quark decay, the signal signature is the presence of a high $p_T$ photon, a b-jet, a muon and missing transverse energy. The signal samples are generated with PROTOS \cite{AguilarSaavedra:2009mx}.

The backgrounds with a prompt photon  arise mainly from  W$\gamma$+jets and Z$\gamma$+jets, and the backgrounds with a fake photon are dominated with  W+jets and $t\bar{t}$ processes. Events with exactly one well isolated photon with $p_T>50$ GeV and muon with $p_T>26$ GeV are selected. Events with two or more b-tagged jets are vetoed to reduce the $t\bar{t}$ contributions. Finally a top mass window is defined to reduce the contribution of SM backgrounds without top quark.

A boosted decision tree (BDT) is constructed to separate signal from SM background. The contribution of the W+jets and W$\gamma$+jets backgrounds is estimated from data and other from simulation. The data is well described by SM prediction and there is no evidence for the signal. Upper limit on Br($t\rightarrow u\gamma$) is 0.0161\%.
\section{Searches for t-q-gluon anomalous interactions}
\begin{wrapfigure}{r}{0.4\textwidth} 
  \begin{center}
    \includegraphics[width=0.4\textwidth]{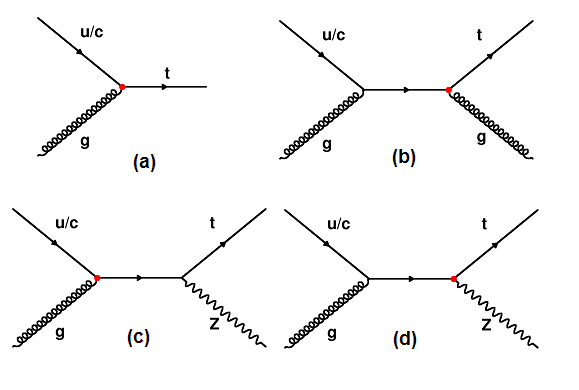}
    \begin{minipage}[b]{14pc}\caption{\label{feynman} Representative Feynman diagrams for FCNC tqg (a,b and c) and tqZ (d) processes.}
\end{minipage}
  \end{center}
\end{wrapfigure} 
Among FCNC top quark decays, $t\rightarrow qg$ is very difficult to distinguish from generic multijet production via quantum chromodynamics (QCD). 
It has therefore been
suggested to search for FCNC couplings in anomalous single top-quark production.

The anomalous $tqg$ interactions can induce various rare processes at hadron colliders. 
The CDF and ATLAS collaborations have chosen the production of a single top quark without any additional particle \cite{Aaltonen:2008qr,TheATLAScollaboration:2013vha}. 
In figure \ref{feynman} (a), the related Feynman diagram is shown. 
The  signature of $qg \rightarrow t \rightarrow lb \nu$ process is different from SM processes. First, W and b from top decay tend to be back to back in the azimuthal plane.
Secondly, W is boosted and its decay products have smaller opening angle. 
Lastly, top is produced three times more than antitop.

Signal events are generated with TOPREX \cite{Slabospitsky:2002ag}, PROTOS \cite{AguilarSaavedra:2009mx} and METOP \cite{Coimbra:2012ys} event generators for the CDF, ATLAS 7 TeV and ATLAS 8 TeV analysis respectively. 
For an efficient background rejection which are dominated by W+jets and QCD multijets, Bayesian Neural-Network (BNN) is used by both collaborations. 
No evidence for such a process is observed.
A binned maximum likelihood fit of the BNN-output distribution is performed to set an upper limit on the FCNC single top quark production cross section. 

The D0 and CMS experiments searched for FCNC production of single top quark and a light quark or gluon, a topology similar to SM t-channel single top quark production. 
A representative Feynman  diagram is shown in figure \ref{feynman} (b). 
The single top quark final state without extra jets that was explored by the CDF and ATLAS Collaborations is not 
\begin{wraptable}{r}{0.62\textwidth}
\caption{\label{summaryreport} The most stringent experimental upper bounds on the top quark FCNC branching ratios at 95\% CL obtained in CDF, D0, ATLAS and CMS from different channels.} 
\begin{tabular}{lllclll}
\br
EXP & $\sqrt{s}$TeV & $\mathcal{L}(fb^{-1})$&Br  & (q=u)\% & (q=c)\% & Ref\\
\mr
ATLAS &7$\&$8  & 25 &\parbox[t]{1mm}{\multirow{3}{*}{\rotatebox[origin=c]{90}{$t\rightarrow qH$ }}} & \multicolumn{2}{c}{0.79 }  & \cite{atlasTQH} \\
CMS &8  & 19.5  & &\multicolumn{2}{c}{0.56 }  & \cite{CMSTQH} \\
&&&&&\\
\mr
CDF &1.8    & 0.11 &\parbox[t]{1mm}{\multirow{3}{*}{\rotatebox[origin=c]{90}{$t\rightarrow q\gamma $ }}}  & \multicolumn{2}{c}{3.2 }  & \cite{Abe:1997fz} \\
CMS &8  & 19.1 &  &0.0161  & 0.182 & \cite{CMS:2014hwa} \\
&&&&&\\
\mr
CDF &1.96   & 2.2  & \parbox[t]{1mm}{\multirow{5}{*}{\rotatebox[origin=c]{90}{$t\rightarrow qg$ }}}   &0.039  & 0.57 & \cite{Aaltonen:2008qr} \\
D0 &1.96    & 2.3  & &0.02  & 0.39 & \cite{Abazov:2010qk} \\
CMS &7  & 4.9 &  &0.56  & 7.12& \cite{CMS:2013nea} \\
CMS &7  & 4.9 & &0.035  & 0.34 & \cite{CMS:2014ffa} \\
ATLAS &8  & 14.2  & &0.0031  & 0.016 & \cite{TheATLAScollaboration:2013vha} \\
\mr
CDF &1.96    & 1.9 &\parbox[t]{1mm}{\multirow{5}{*}{\rotatebox[origin=c]{90}{$t\rightarrow qZ$ }}} & \multicolumn{2}{c}{3.7 }  & \cite{Aaltonen:2008ac} \\
D0 &1.96    & 4.1 &  & \multicolumn{2}{c}{3.2 }  & \cite{Abazov:2010qk} \\
CMS &7  & 4.9 &&0.51  & 11.40 & \cite{CMS:2013nea} \\
ATLAS &7    & 2.1&& \multicolumn{2}{c}{0.73 }  & \cite{Aad:2012ij} \\
CMS &7$\&$8   & 24.7 &  & \multicolumn{2}{c}{0.05 }  & \cite{Chatrchyan:2013nwa} \\
\br
\end{tabular}
\end{wraptable} 
considered due to its different final state topology and significantly smaller signal event yield. 
The presence of exactly one isolated charged lepton (e or $\mu$ for D0 and $\mu$ for CMS),  one light-flavour jet in the forward region,
one b-jet and significant missing transverse momentum due to the presence of a neutrino are the signatures of this final state. 
The main backgrounds are from W+jets production, including W+c-quarks and W+b-quarks. The QCD can also contribute in final selection with a fake lepton. To suppress QCD background, a dedicated BNN is used in CMS analysis. 
Both experiments have trained a BNN for the discrimination of FCNC signal from the SM processes. Since the data are consistent with the background expectation, 
the 95\% upper limits is set on the FCNC cross sections and couplings using BNN
distributions.

The obtained upper limits with the experimental conditions are summarised in table \ref{summaryreport}. 
The limits on the $BR(t\rightarrow u(c)g)<0.0031 (0.016)\%$ obtained by ATLAS collaboration is the most stringent limits to date.

\section{Searches for t-q-Z anomalous interactions}
The D0 \cite{Abazov:2010qk}, CMS \cite{Chatrchyan:2013nwa} and ATLAS \cite{Aad:2012ij} collaborations made searches for $tqZ$ FCNC interactions in $t\bar{t} $ events in which either the top or antitop quark has decayed into a Z boson and
a quark, while the remaining top or antitop quark decayed to Wb. Only leptonic decays of the Z and W bosons were considered, yielding trilepton final states. This mode provides a distinct signature with low SM background, albeit suffers from statistical uncertainties. The backgrounds include events
with three real final-state charged leptons mainly from WZ and ZZ processes, as well as events with at least one fake leptons dominated by Z+jets and $t\bar{t} $ processes.
A similar search is done by CDF collaboration \cite{Aaltonen:2008ac} while the hadronic decays of the Z boson is considered. The signal events are generated with Pythia \cite{Sjostrand:2000wi} in CDF and D0 analysis and TopRex \cite{Slabospitsky:2002ag} and MadGraph \cite{Maltoni:2002qb} for ATLAS and CMS analysis respectively.
The data are consistent with the SM background prediction, and upper limit on the branching fraction $BR(t\rightarrow qZ)$ are set which are reported in table \ref{summaryreport}.

The signature of a single top quark production in association with a Z boson is used by CMS collaboration to search for $tqZ$ and $tqg$ anomalous interactions simultaneously, via diagrams (c) and (d) in figure \ref{feynman} \cite{CMS:2013nea}. The signal events are discriminated from SM background events using BDT. No FCNC signal is observed and the upper bounds on the  anomalous branching 
\begin{wrapfigure}{r}{0.42\textwidth} 
    \includegraphics[width=0.45\textwidth]{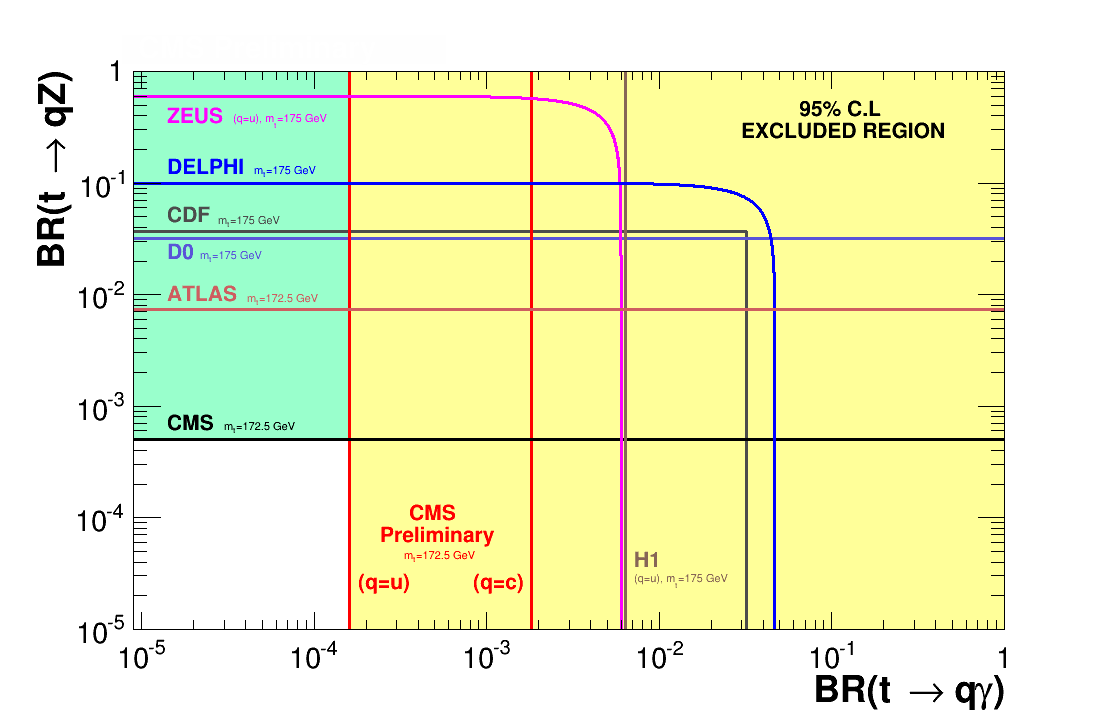}
    \begin{minipage}[b]{16pc}\caption{\label{2Dplot} The observed 95\% C.L. upper limit on the $BR(t\rightarrow q\gamma)$ vs $BR(t\rightarrow qZ)$ for the DELPHI, ZEUS, H1, D0, CDF, ATLAS and CMS collaborations.}
\end{minipage}
   \end{wrapfigure} 
fractions are derived  which is shown in table \ref{summaryreport}.
The $tqg$ upper limits provide an interesting cross-check 
in a different physics process
and the $tqZ$ upper limits can be combined with other searches to improve the current limits. 
\section{Summary}
The anomalous production and decay of top quark are two complementary channels which have been considered by D0, CDF, ATLAS and CMS collaborations in order to search for anomalous FCNC top couplings. No excess over the SM expectation is found and the strength of the anomalous couplings are limited consequently. Although the SM predicts top quark anomalous branching ratios many order of magnitude below the current experimental limits, experiments are closing to the regions which are predicted by some beyond SM models as shown in figure \ref{2Dplot} in the $BR(t\rightarrow q\gamma)$ vs $BR(t\rightarrow qZ)$ plane.


\section*{References}
\begin{thebibliography}{}
\bibitem{gim}
 S.~L.~Glashow, J.~Iliopoulos and L.~Maiani,
Phys.\ Rev.\ D {\bf 2}, 1285 (1970). 

\bibitem{smbr}
J.~A.~Aguilar-Saavedra,
Acta Phys.\ Polon.\ B {\bf 35}, 2695 (2004)
[hep-ph/0409342]. 

\bibitem{Chatrchyan:2008aa} 
CMS Collaboration,
  JINST {\bf 3}, S08004 (2008).

\bibitem{Aad:2008zzm} 
ATLAS Collaboration,
  JINST {\bf 3}, S08003 (2008).
  
  \bibitem{Acosta:2004yw} 
CDF Collaboration,
  Phys.\ Rev.\ D {\bf 71}, 032001 (2005)
  [hep-ex/0412071].
    
  \bibitem{Abazov:2005pn} 
 D0 Collaboration,
  Nucl.\ Instrum.\ Meth.\ A {\bf 565}, 463 (2006)
  [physics/0507191 [physics.ins-det]].

\bibitem{Aad:2012higgs}
ATLAS Collaboration,
Phys.\ Lett.\ B {\bf 716}, 2 (2012)

\bibitem{Chatrchyan:2012higgs} 
CMS Collaboration,
Phys.\ Lett.\ B {\bf 716}, 30 (2012)

\bibitem{atlasTQH} 
  ATLAS Collaboration,
  JHEP {\bf 1406}, 008 (2014)
  [arXiv:1403.6293 [hep-ex]].

\bibitem{CMSTQH} 
CMS Collaboration,
  CMS-PAS-HIG-13-034.

\bibitem{AguilarSaavedra:2009mx} 
  J.~A.~Aguilar-Saavedra,
  Nucl.\ Phys.\ B {\bf 821}, 215 (2009)
  [arXiv:0904.2387 [hep-ph]].

\bibitem{Chatrchyan:2014aea} 
 CMS Collaboration,
  Phys.\ Rev.\ D {\bf 90}, 032006 (2014)
  [arXiv:1404.5801 [hep-ex]].

\bibitem{CMS:2013eua} 
CMS Collaboration,
  CMS-PAS-HIG-13-025.

  
\bibitem{Maltoni:2002qb} 
  F.~Maltoni and T.~Stelzer,
  JHEP {\bf 0302}, 027 (2003)
  [hep-ph/0208156].
  
  \bibitem{Abe:1997fz} 
CDF Collaboration,
  Phys.\ Rev.\ Lett.\  {\bf 80}, 2525 (1998).

\bibitem{Abramowicz:2011tv} 
ZEUS Collaboration,
  Phys.\ Lett.\ B {\bf 708}, 27 (2012)
  [arXiv:1111.3901 [hep-ex]].
  
  \bibitem{Aaron:2009vv} 
H1 Collaboration,
  Phys.\ Lett.\ B {\bf 678}, 450 (2009)
  [arXiv:0904.3876 [hep-ex]].
  
  \bibitem{Abdallah:2003wf} 
DELPHI Collaboration,
  Phys.\ Lett.\ B {\bf 590}, 21 (2004)
  [hep-ex/0404014].
  
  \bibitem{CMS:2014hwa} 
CMS Collaboration,
association with a photon,''
  CMS-PAS-TOP-14-003.
  
  \bibitem{Aaltonen:2008qr} 
CDF Collaboration,
  Phys.\ Rev.\ Lett.\  {\bf 102}, 151801 (2009)
  [arXiv:0812.3400 [hep-ex]].
  
  \bibitem{TheATLAScollaboration:2013vha} 
ATLAS Collaboration,
  ATLAS-CONF-2013-063, ATLAS-COM-CONF-2013-064.

  \bibitem{Slabospitsky:2002ag} 
  S.~R.~Slabospitsky and L.~Sonnenschein,
  Comput.\ Phys.\ Commun.\  {\bf 148}, 87 (2002)
  [hep-ph/0201292].
 
\bibitem{Coimbra:2012ys} 
  R.~Coimbra, A.~Onofre, R.~Santos and M.~Won,
  Eur.\ Phys.\ J.\ C {\bf 72}, 2222 (2012)
  [arXiv:1207.7026 [hep-ph]].
  
  \bibitem{Abazov:2010qk} 
D0 Collaboration,
  Phys.\ Lett.\ B {\bf 693}, 81 (2010)
  [arXiv:1006.3575 [hep-ex]].
  
  \bibitem{CMS:2014ffa} 
CMS Collaboration,
  CMS-PAS-TOP-14-007.
  
  \bibitem{CMS:2013nea} 
  CMS Collaboration,
  CMS-PAS-TOP-12-021.

  \bibitem{Aaltonen:2008ac} 
CDF Collaboration,
  Phys.\ Rev.\ Lett.\  {\bf 101}, 192002 (2008)
  [arXiv:0805.2109 [hep-ex]].
 
  \bibitem{Abazov:2010qk} 
D0 Collaboration,
  Phys.\ Lett.\ B {\bf 693}, 81 (2010)
  [arXiv:1006.3575 [hep-ex]].
  
  \bibitem{Aad:2012ij} 
ATLAS Collaboration,
  JHEP {\bf 1209}, 139 (2012)
  [arXiv:1206.0257 [hep-ex]].
  
  \bibitem{Chatrchyan:2013nwa} 
CMS Collaboration,
  Phys.\ Rev.\ Lett.\  {\bf 112}, 171802 (2014)
  [arXiv:1312.4194 [hep-ex]].
  
  \bibitem{Sjostrand:2000wi} 
  T.~Sjostrand {\it et al.} ,
  Comput.\ Phys.\ Commun.\  {\bf 135}, 238 (2001)
  [hep-ph/0010017].
  
\end{thebibliography}
\end{document}